\documentclass[slac_one]{revtex4}
\usepackage{graphicx}
\usepackage{fancyhdr}
\newcommand{\bd}{B_{d}^{0}}
\newcommand{\jpsi}{J/\psi}
\newcommand{\kstar}{K^{*0}}

\newcommand{\bddecay}{\bd\to\jpsi\kstar}
\newcommand{\bs}{B_{s}^{0}}

\newcommand{\bsdecay}{\bs\to\jpsi\phi}
\newcommand{\acero}{|A_{0}|^{2}}
\newcommand{\all}{|A_{\parallel}|^{2}}
\newcommand{\aperp}{|A_{\perp}|^{2}}
\newcommand{\dll}{\delta_{\parallel}}
\newcommand{\duno}{\delta_{1}}
\newcommand{\ddos}{\delta_{2}}
\newcommand{\taus}{\bar{\tau}_{s}}
\newcommand{\taud}{\tau_{d}}
\newcommand{\tstd}{\taus/\taud}

\newcommand{\ds}{\delta_{s}}

\begin{document}

\title{Time-dependent angular analysis of the decays {\boldmath $\bddecay$ \unboldmath} and 
       {\boldmath $\bsdecay$\unboldmath}}

\author{G.A.~Garcia-Guerra (for the D0 Collaboration)}
\affiliation{Departamento de F\'isica, CINVESTAV, AP 14-740, 07000 Mexico DF, Mexico}

\begin{abstract}
In this paper we present the description of the flavor-untagged decays $\bddecay$ 
and $\bsdecay$ in the transversity basis.  The study of these $B$ mesons in that 
basis makes it possible to extract information about flavor SU(3) symmetry and to 
verify if the factorization assumption is feasible for the decay $\bddecay$. 
The lifetime ratio $\tstd$ is also extracted with this description.
\end{abstract}

\maketitle

\thispagestyle{fancy}

\section{INTRODUCTION}
Both decays considered in the present analysis, $\bddecay$ and 
$\bsdecay$, are decays of a pseudo-scalar to a vector-vector intermediate state.
The observables of the angular distributions, the linear polarization amplitudes
and the strong relative phases of the $B$ mesons that decay in such a way, can 
be extracted by their description in the transversity basis~\cite{fleischer}. By
measuring those observables, we can obtain important information about flavor 
SU(3) symmetry and the factorization assumption related with these decays. The 
former requires that the linear polarization amplitudes and the strong relative 
phases characterizing these decays should have the same 
values~\cite{fleischer,gronau}\footnote{An interesting discussion in Ref.~\cite{gronau} 
states that the flavor symmetry comes from U(3) rather than SU(3).}. 
Factorization states that, in the absence of final-state interactions (FSI), the 
strong phases are 0 (mod\thinspace$\pi$)~\cite{fleischer,browder} for the 
$\bddecay$. In this paper we describe the flavor-untagged\footnote{By not identifying the 
initial $B$ meson flavor} decays $\bddecay$ and $\bsdecay$ in the transversity 
basis. The final measurements of these analyses are reported in Ref.~\cite{bd.bs.untagged}.

\section{THE ANGULAR DESCRIPTION OF THE DECAYS $\bddecay$ AND $\bsdecay$}
\begin{figure} [h]
\includegraphics[height=0.25\textwidth]{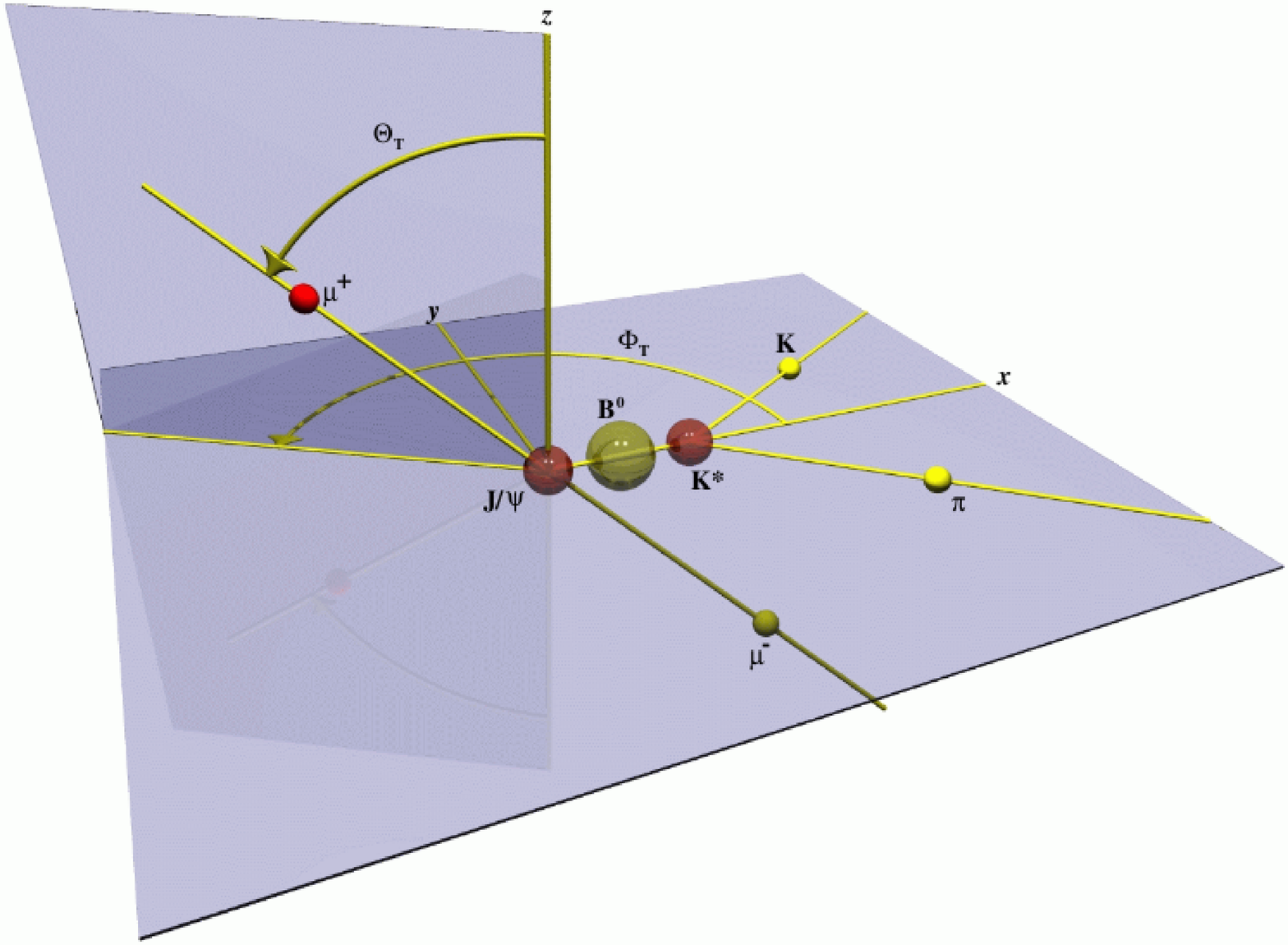}\hspace{1cm}
\includegraphics[height=0.25\textwidth]{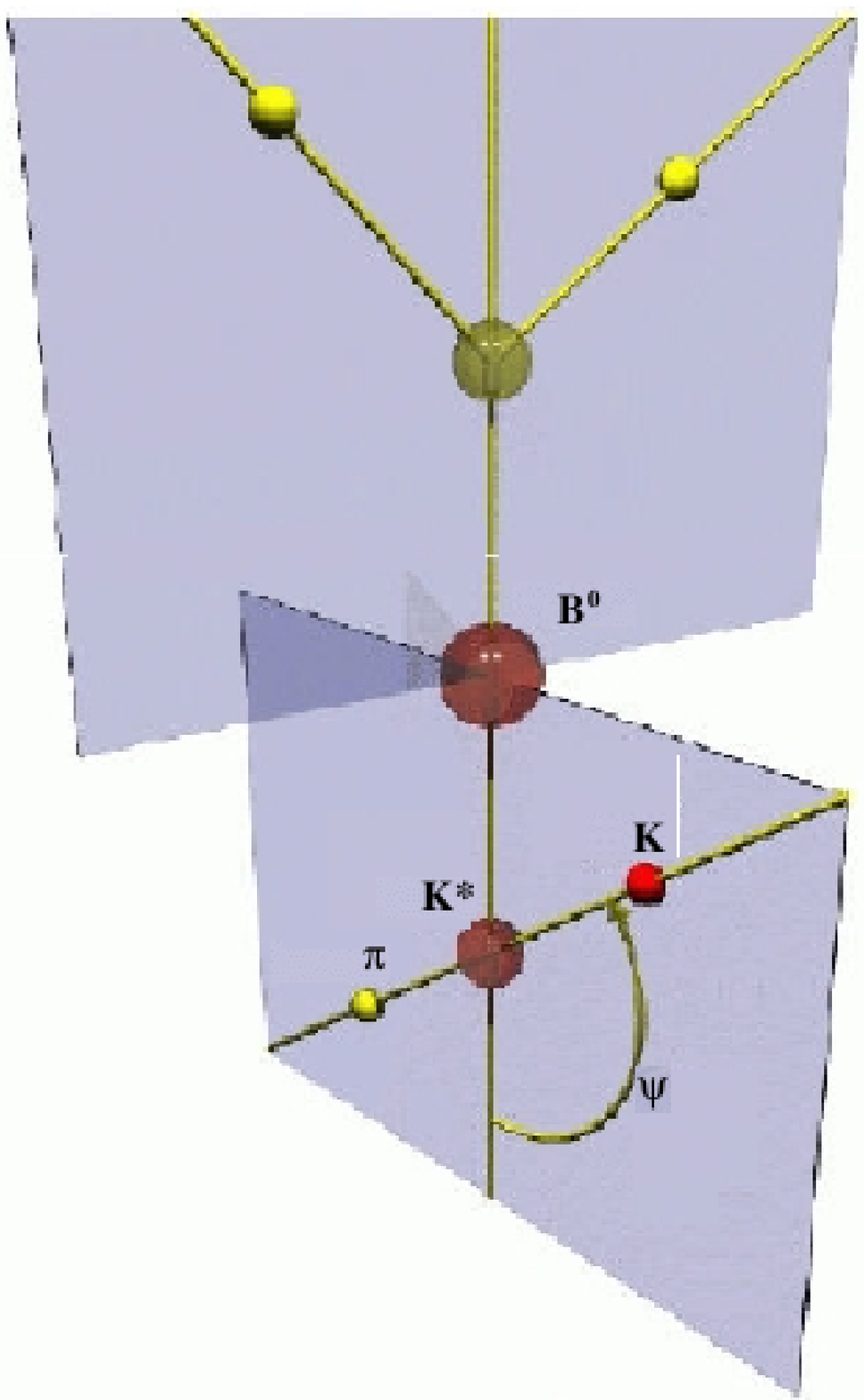}\hspace{2cm}
\includegraphics[height=0.25\textwidth]{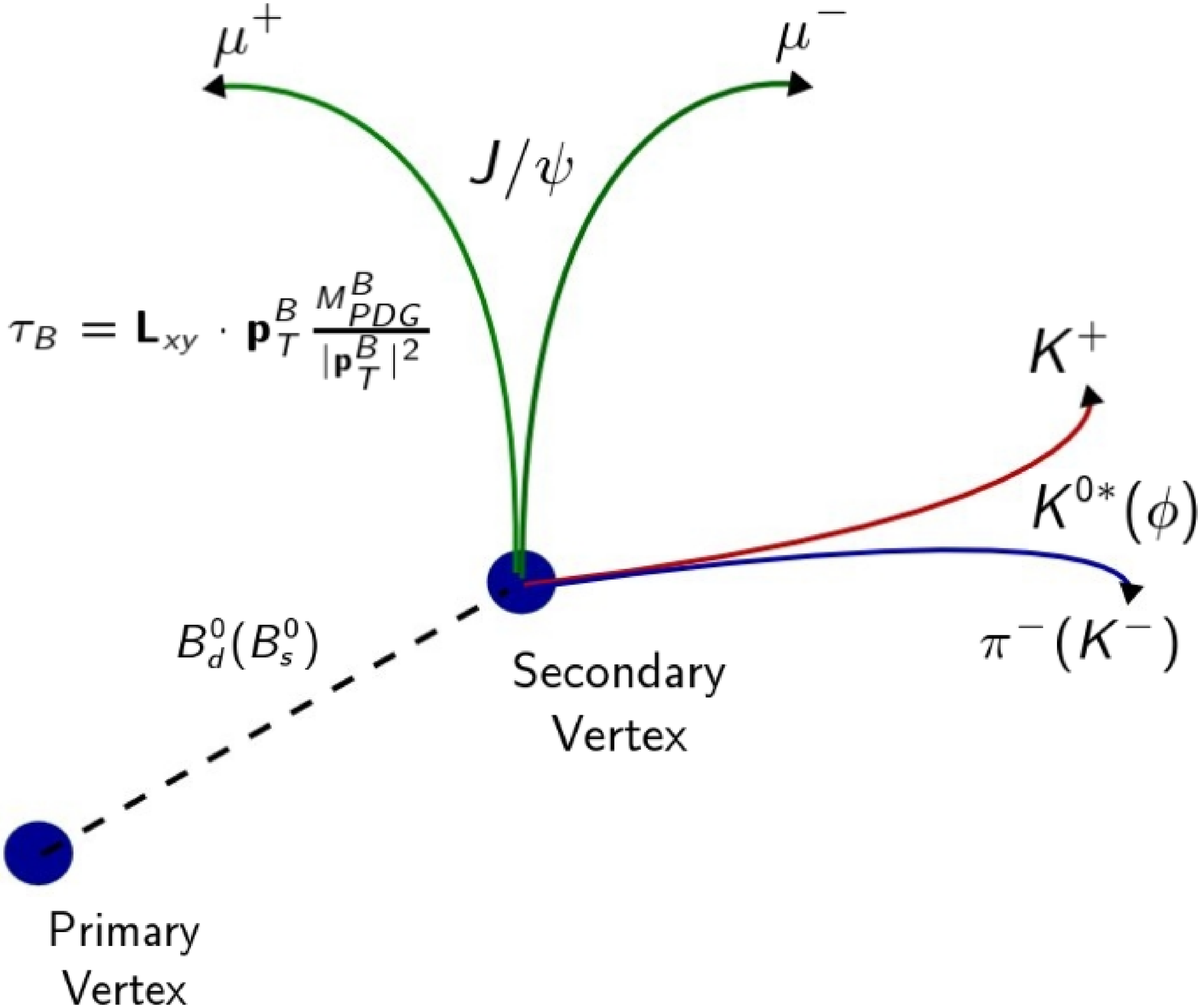}
\caption{The angular variables of the transversity basis $\varphi$ and $\cos\theta$
         are defined in the $\jpsi$ rest frame (left), and $\cos\theta$ in the
         $\kstar$ rest frame relative to the negative direction of the $\jpsi$ in
         that frame (center). The translation of this basis to the decay 
         $\bsdecay$ is straightforward due to the same 4-vertex track topology 
         for the decays (right).\label{fig:topology.transv}}
\end{figure}

Due to the same 4-track topology, the decays under study can be described by the
same transversity basis (see Fig.~\ref{fig:topology.transv}). 
We denote by $\boldmath{\omega}=\left\{\varphi,\cos\theta,\cos\psi\right\}$ the 
set of the angular variables for this basis.

For the $\bd$ system, we take into account the interference between the $K\pi$ 
$P$- and $S$-wave amplitudes as described in Ref.~\cite{babar}.  Therefore, the 
differential decay rate for the untagged decay $\bddecay$ is given 
by~\cite{fleischer,babar}:
\begin{eqnarray}\label{decay.distrib.bd}
\nonumber\frac{d^{4}{\cal P}}{d\boldmath{\omega}\;dt}& \propto & e^{-t/\taud}\left\{\cos^{2}\lambda\left[\acero f_{1}(\boldmath{\omega})
+\all f_{2}(\boldmath{\omega})+\aperp f_{3}(\boldmath{\omega})-\zeta\;\mathrm{Im}(A_{\parallel}^{*}A_{\perp})f_{4}(\boldmath{\omega})
+\mathrm{Re}(A_{0}^{*}A_{\parallel})f_{5}(\boldmath{\omega})\right.\right.\\
&+&\left.\zeta\;\mathrm{Im}(A_{0}^{*}A_{\perp})f_{6}(\boldmath{\omega})\right]
+\sin^{2}\lambda\cdot f_{7}(\boldmath{\omega})
+\frac{1}{2}\sin2\lambda\left[f_{8}(\boldmath{\omega})\cos\left(\ddos - \duno - \ds\right)|A_{\parallel}|\right.\\
&+&\nonumber\left.f_{9}(\boldmath{\omega})\sin\left(\ddos - \ds\right)|A_{\perp}|
+\left.f_{10}(\boldmath{\omega})\cos\ds\cdot|A_{0}|\right]\right\},
\end{eqnarray}
where $\taud$ is the $\bd$ lifetime, $\zeta=+1(\zeta=-1)$ is for an initially 
produced $K^{+}(K^{-})$; $\lambda,\ds$, and $f_{i}(\boldmath{\omega})$ are 
defined in Refs.~\cite{fleischer,babar}. If any of the strong phases, 
$\duno\equiv\arg[A_{\parallel}^{*}A_{\perp}]$ and $\ddos\equiv\arg[A_{0}^{*}A_{\perp}]$, 
are not consistent with $0 ($mod$\thinspace\pi)$, then the factorization assumption 
is not valid for the decay $\bddecay$.

In the $\bs$ system, since the standard model predicts a very small CP-violating 
phase~\cite{lenz}, we assume CP conservation for simplicity. From this, the 
differential decay rate for the untagged decay $\bsdecay$ is given by~\cite{fleischer} :
\begin{equation}\label{decay.distrib.bs}
 d^{4}{\cal P}/\left(d\omega\;dt\right)\propto e^{-\Gamma_{L}t}\left[\acero f_{1}(\omega)+\all f_{2}(\omega) 
 +\mathrm{Re}(A_{0}^{*}A_{\parallel})f_{5}(\omega)\right] + e^{-\Gamma_{H}t}|A_{\perp}|^{2}f_{3}(\omega),
\end{equation}
where $\Gamma_{L(H)}\equiv1/\tau_{L(H)}$ is the
inverse of the lifetime corresponding to the light (heavy) mass eigenstate. For
this decay, we have access to the same linear polarization amplitudes as for the
$\bd$ and the phase $\dll$, which is related with $\duno$ and $\ddos$ by means of
the relation $\dll=\ddos-\duno$. In this analysis, we also measure the mean 
lifetime $\taus\equiv1/\bar{\Gamma}=2/\left(\Gamma_{L}+\Gamma_{H}\right)$.
If flavor SU(3)~\cite{fleischer} (or U(3)~\cite{gronau}) symmetry is valid, then
the linear amplitudes and the strong phases should be consistent with being equal
for both the $\bd$ and $\bs$ mesons.

\section{THE MONTE CARLO REWEIGHTING\label{section:mc.r}}
\begin{figure} [h]
\begin{tabular}{cc}
\includegraphics[height=0.19\textwidth]{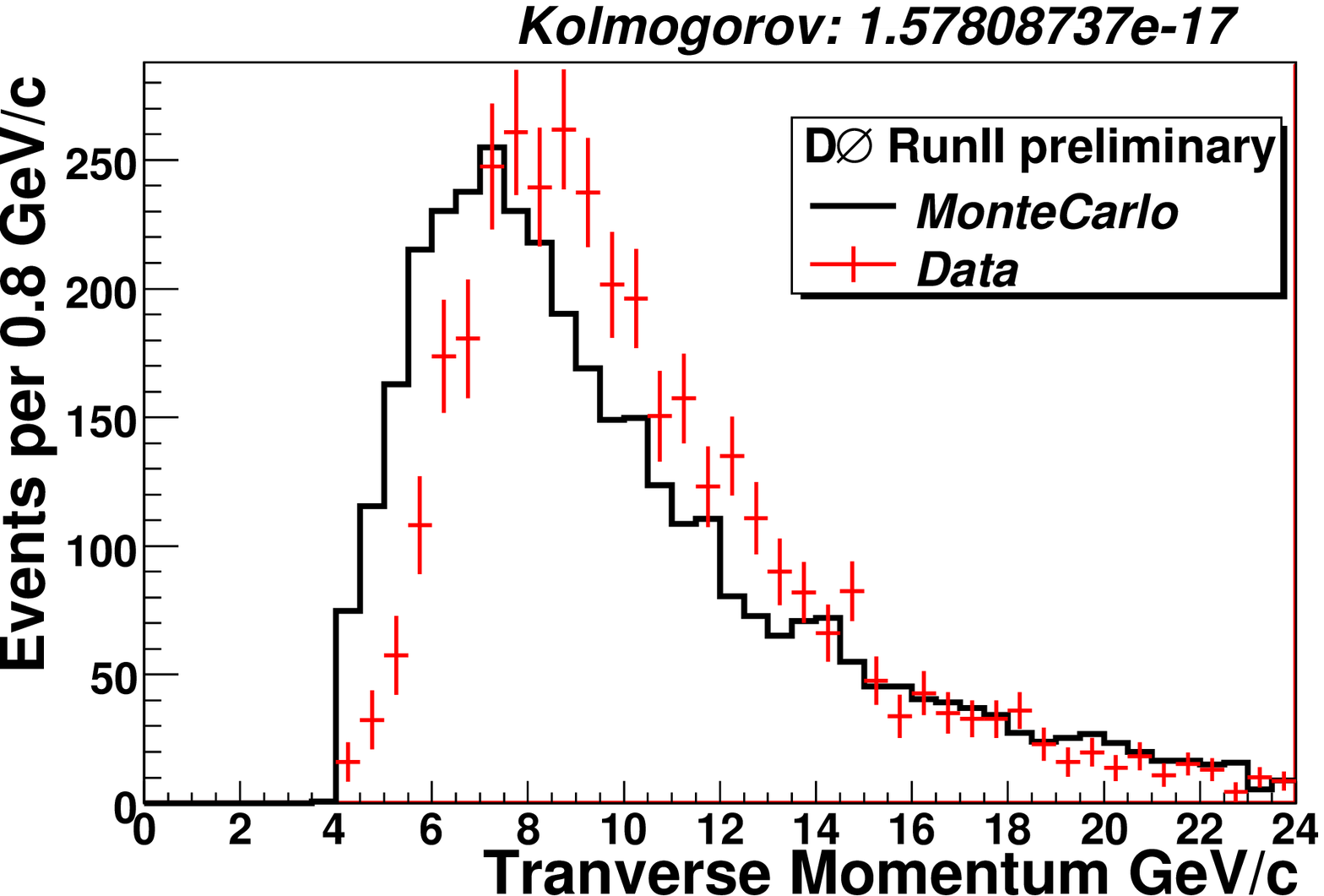}&\includegraphics[height=0.19\textwidth]{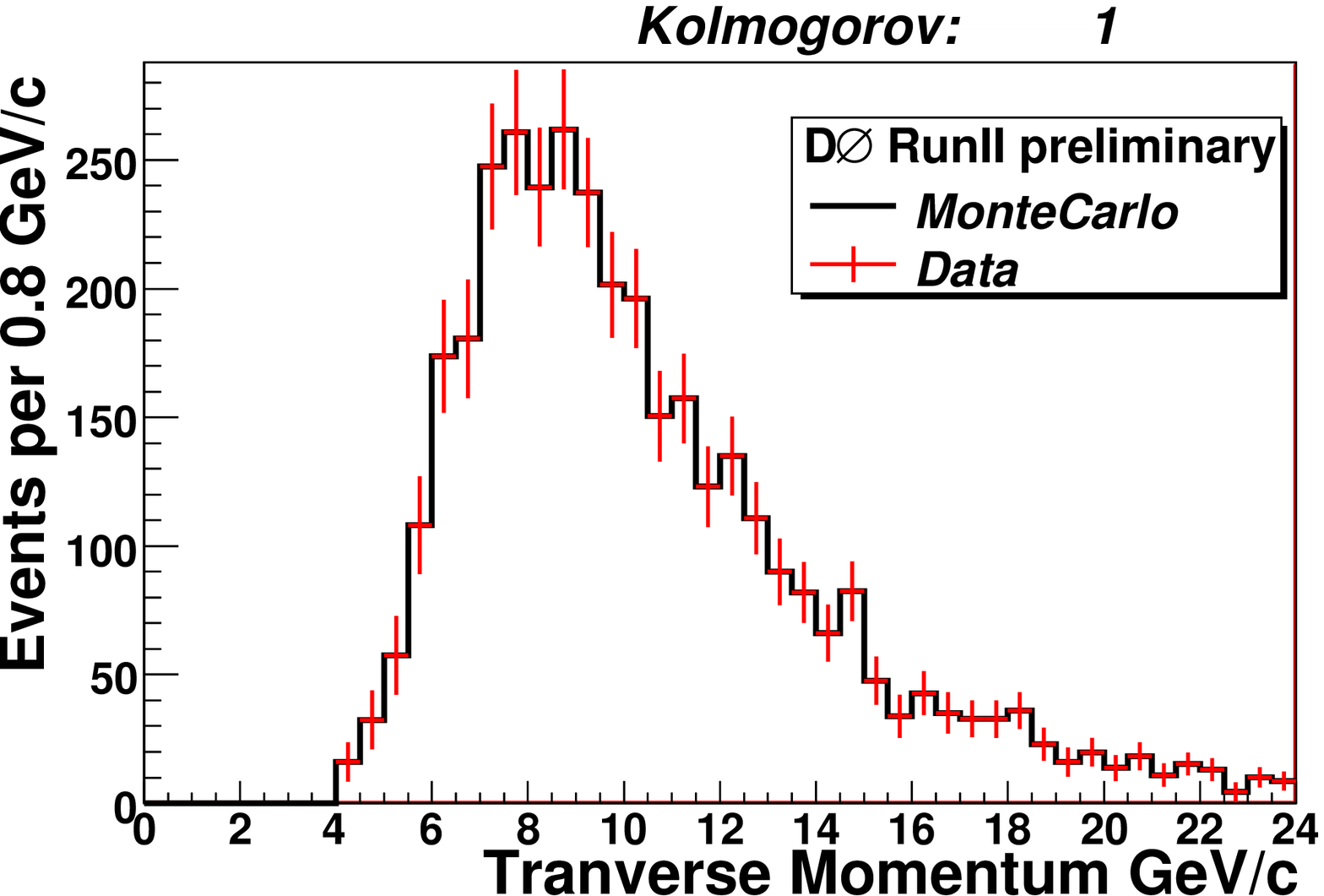}\\
\includegraphics[height=0.19\textwidth]{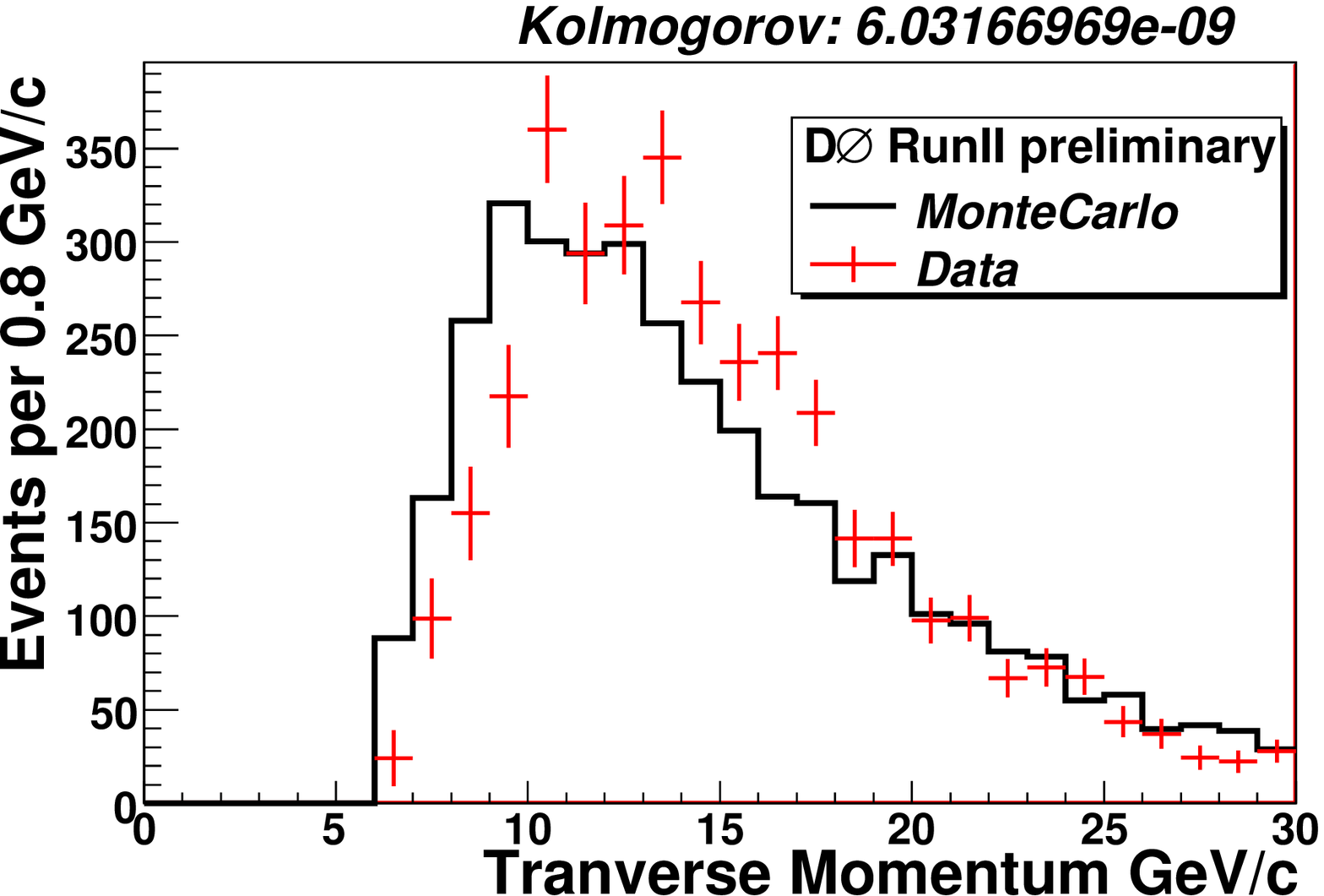}&\includegraphics[height=0.19\textwidth]{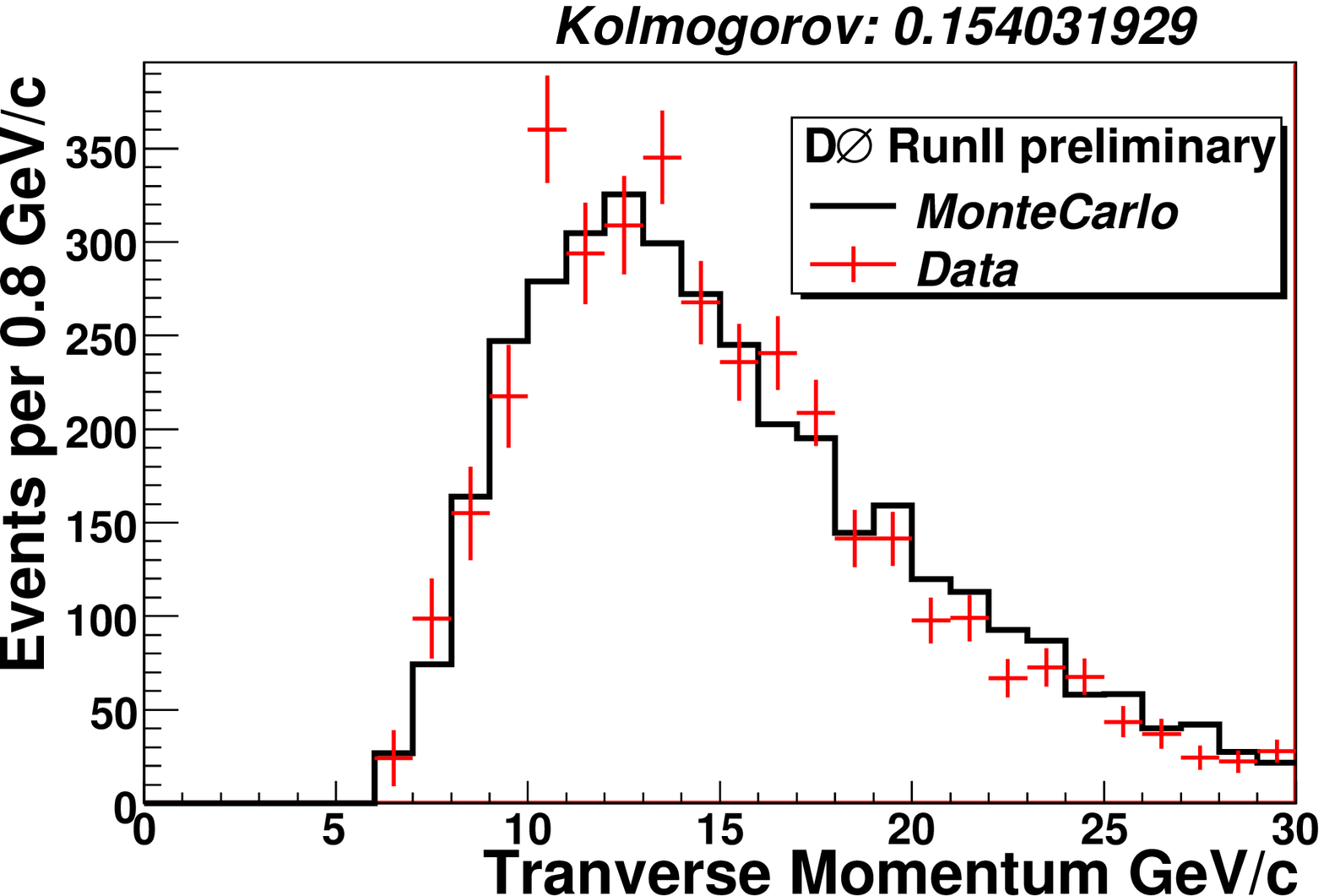}
\end{tabular}
\caption{$p_{T}(\jpsi)$ (top) and $p_{T}(\bd)$ (bottom) distributions before 
         (left) and after (right) reweighting for the decay $\bddecay$ using the 
         histogram of the $p_{T}(\jpsi)$ distribution. The reweighting method
         improves the MC distributions, as can be seen from the Kolmogorov tests.
         Similar histograms are obtained for the decay $\bsdecay$\label{fig:mc.r}}
\end{figure}

The distributions of certain kinematic variables in the Monte Carlo (MC) 
simulation, such as the transverse momentum $p_{T}$ of the particles, do not 
agree well with data. The D0 Detector~\cite{run2det} has a tracking and a muon 
systems such that the muon reconstruction is well understood. Because of this, we 
choose the $p_{T}(\jpsi)$ distribution to weight the generated MC distributions 
to agree with that of data. The comparison of some kinematic distributions before 
and after reweighting is shown in Fig.~\ref{fig:mc.r}.

\section{THE ANGULAR EFFICIENCY $\epsilon({\boldmath\omega})$}
\begin{figure}[h]
\includegraphics[height=0.23\textwidth]{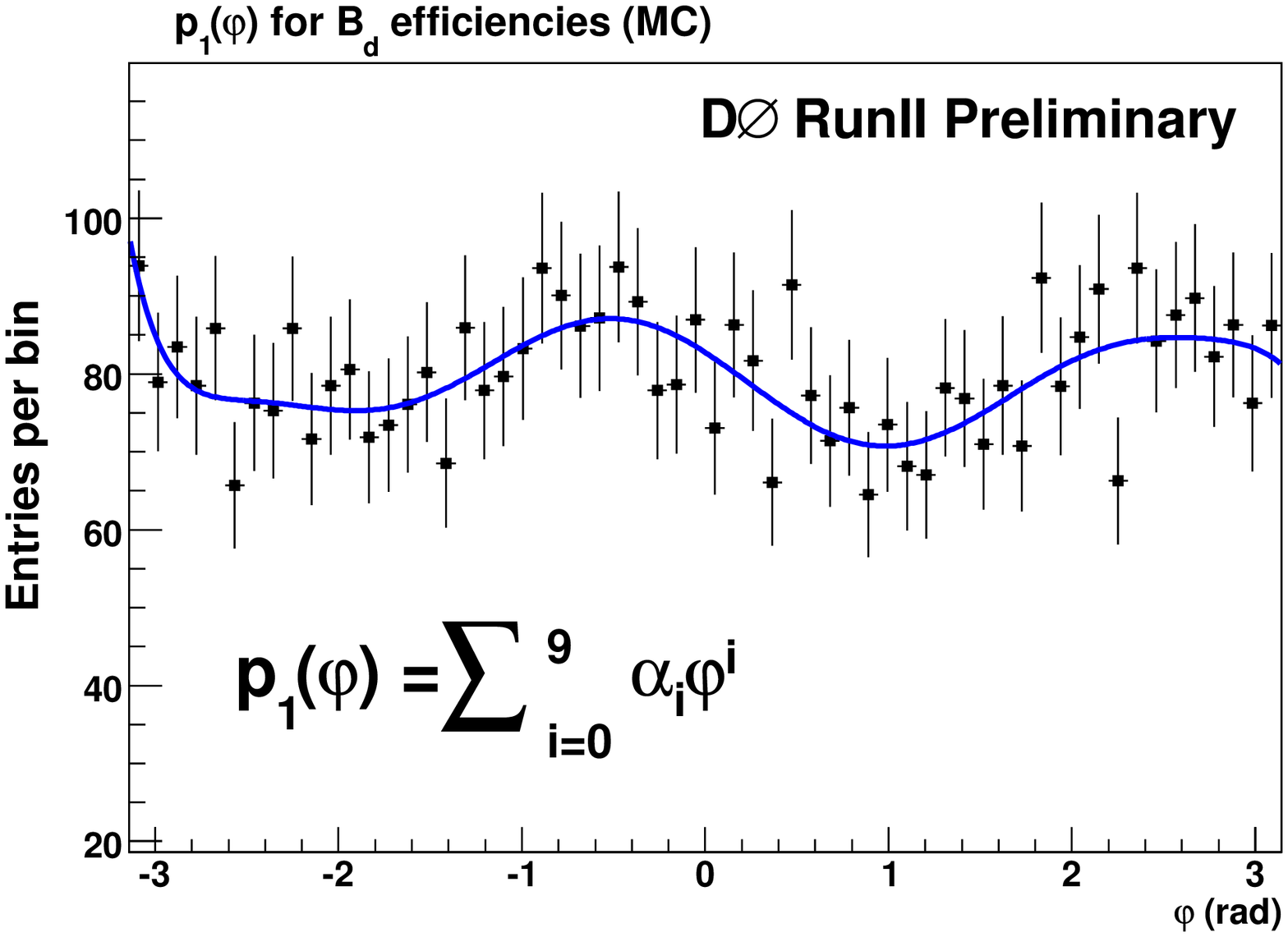}
\includegraphics[height=0.235\textwidth]{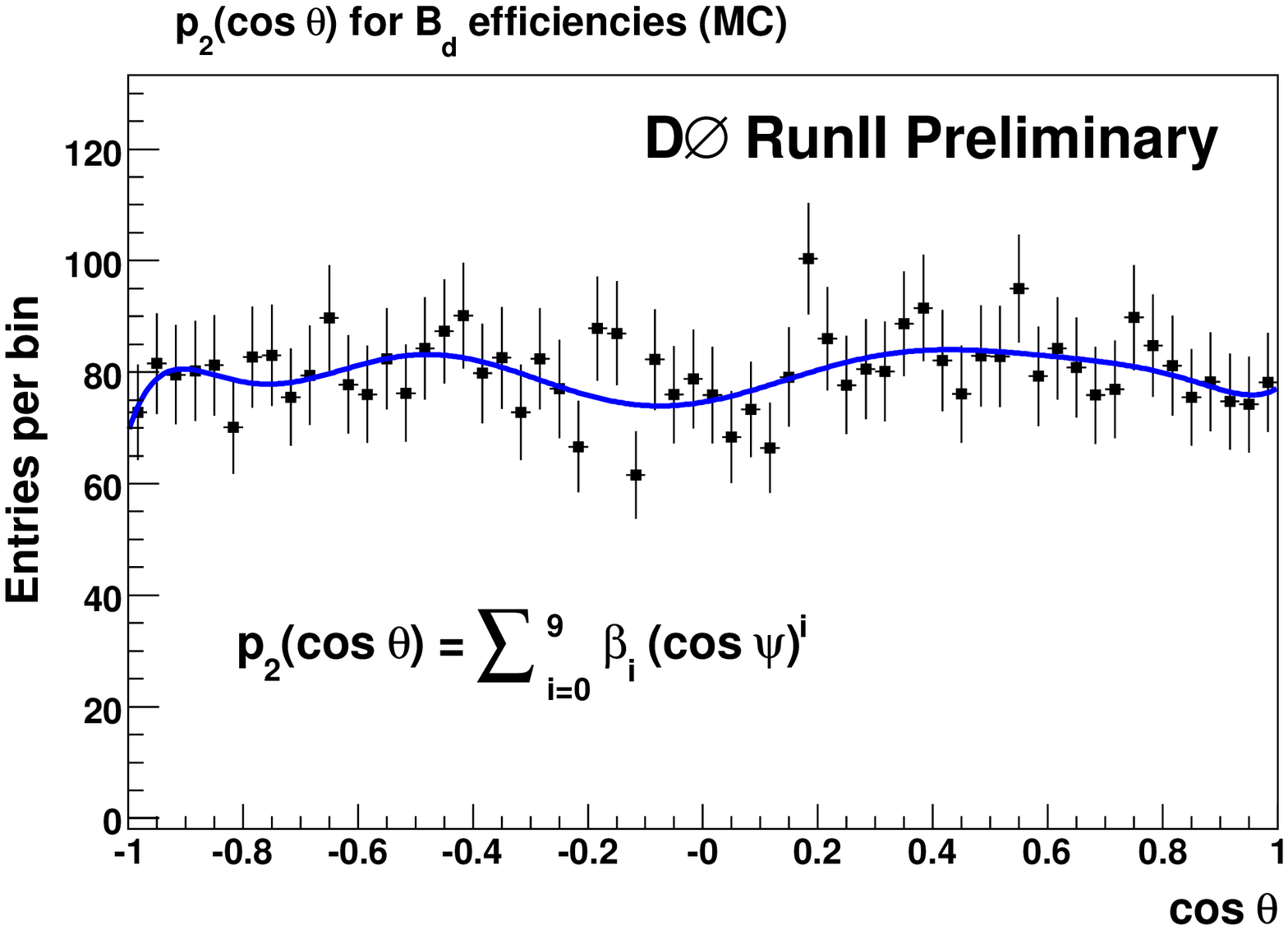}
\includegraphics[height=0.23\textwidth]{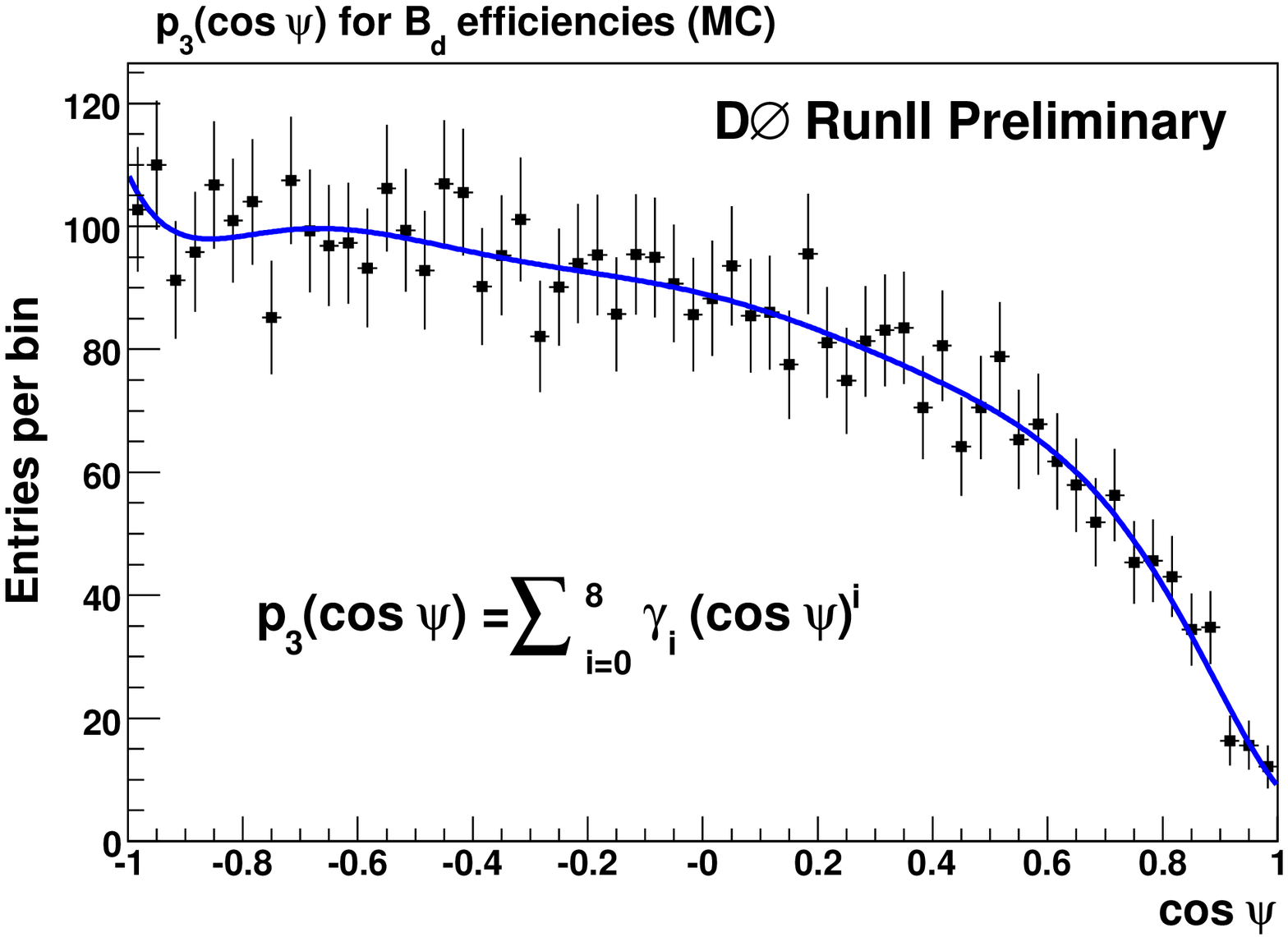}
\caption{Polynomials $p_{i}$ for the $\bd$ angular efficiencies.\label{fig:pols.bd}}
\end{figure}

We need to take into account the effect of the detector in the theoretical 
distributions Eqs. (\ref{decay.distrib.bd}) and (\ref{decay.distrib.bs}). We model this effect 
assuming that it can be written as a product of three polynomials 
$\epsilon(\omega)=p_{1}(\varphi)p_{2}(\cos\theta)p_{3}(\cos\psi)$. The 
coefficients of these polynomials are obtained by fitting the angular distributions
of the reweighted MC.  For the $\bd$ decays, the polynomials that give us the 
best modeling for $\epsilon(\omega)$ are those shown in Fig.~\ref{fig:pols.bd}.
We follow the same procedure to obtain the polynomials for the decay $\bsdecay$.

\section{THE ANGULAR PROBABILITY DISTRIBUTION FUNCTIONS\label{section:pdfs}}
To extract the angular and lifetime parameters that describe the flavor-untagged decays $\bddecay$ and $\bsdecay$, we 
need to write the Eqs. (\ref{decay.distrib.bd}) and (\ref{decay.distrib.bs}) as 
probability distribution functions (pdf).  The angular pdfs for the $j$-th 
$\bd$ and $\bs$ candidates are given by:
\begin{eqnarray}\label{funcionangularsignalB0d}
\mathcal{F}_{\bd}(\acero,\all,\duno,\ddos,\taud)&=&\frac{\varepsilon(\omega_{j})}{N_{\bd,sig}}\sum_{i=1}^{10}g_{i}f_{i}(\omega_{j}),\\
\mathcal{F}_{\bs}(\acero,\all,\dll,\taus)&=&\frac{\epsilon_{s}(\omega_{j})}{N_{\bs,sig}}\left[\mathcal{T}_{sig}^{L}\sum_{i=1,2,5}h_{i}f_{i}(\omega_{j})+\mathcal{T}_{sig}^{H}h_{3}f_{3}(\omega_{j})\right],
\end{eqnarray}
respectively, where $g_{i}(h_{i})$ are the coefficients of Eq.(\ref{decay.distrib.bd}) 
[Eq.(\ref{decay.distrib.bs})] and $N_{B_{d(s)},sig}$ is the normalization factor.  
The complete description of the log-likehood functions for both decays is reported 
in Ref.~\cite{bd.bs.untagged}.

\begin{acknowledgments}
Work supported by CINVESTAV-Mexico and Consejo Nacional de Ciencia y 
Tecnolog\'ia (CONACyT).
\end{acknowledgments}

\end{document}